\documentstyle[eqsecnum,aps]{revtex}

\begin{document}
\draft
\title{Absence of a Zero Temperature Vortex Solid Phase in Strongly Disordered Superconducting Bi Films}
\author{J. A. Chervenak* and J. M. Valles, Jr.}
\address{
Department of Physics, Brown University, Providence, RI  02912
}
\date{\today}
\maketitle
\begin{abstract}
We present low temperature measurements of the resistance in magnetic field of superconducting ultrathin amorphous Bi films with normal state sheet resistances, $R_N$, near the resistance quantum, $R_Q={\hbar\over {e^2}}$.  For $R_N<R_Q$, the tails of the resistive transitions show the thermally activated flux flow signature characteristic of defect motion in a vortex solid with a finite correlation length. When $R_N$ exceeds $R_Q$, the tails become non-activated.  We conclude that in films where $R_N>R_Q$ there is no vortex solid and, hence, no zero resistance state in magnetic field. We describe how disorder induced quantum and/or mesoscopic fluctuations can eliminate the vortex solid and also discuss implications for the magnetic-field-tuned superconductor-insulator transition.
\end{abstract}
\twocolumn
\narrowtext

The question of whether a superconducting film can superconduct in the presence of disorder and a perpendicular magnetic field continues to receive a great deal of theoretical and experimental attention\cite{dsfish,giam,blatter,ghosal,shim,ephron,kes1}.  As a general question, it reflects on investigations of how disorder influences both thermal and quantum phase transitions\cite{fisher,belkirk}.   More specifically, the answer impacts our understanding of the possible electronic phases of strongly correlated electronic systems\cite{sitreviews,sondhi}.  For the zero dc resistance state of a type II superconductor to persist in the mixed state, the magnetic field induced flux lines must remain fixed in the presence of a vanishingly small applied current.  Disorder pins flux lines and opposes the tendency of the flux line system to form an ordered solid phase\cite{larkin}.  Moreover, the disorder, when it is very strong, can create mesoscopic and quantum fluctuation effects that may also oppose ordering \cite{blatter,zhou}.  

Experiments suggest that flux line solids with finite correlation lengths do form in low sheet resistance, $R_N<<R_Q=\hbar/e^2=4.12 k\Omega$ (i.e. weakly disordered) quasi-two-dimensional superconductors.  Signatures of melting \cite{kes} and thermally activated flux flow (TAFF) resistance with an activation energy consistent with the activation of mobile defects in a solid phase have been observed\cite{feigelman,TAFF1,chervenak1}.  While these films are not expected to superconduct at nonzero temperatures because the energy to activate defects (e.g., edge dislocation pairs) is finite in two dimensions, it is possible that they superconduct at zero temperature.  

Recently, a great deal of work has focused on the strongly disordered regime where $R_N\rightarrow R_Q$, and the superfluid density is low\cite{liu,hsu,chervenak2}.  There have been opposing opinions over the extent of a vortex solid phase.  Magnetotransport measurements on numerous systems have been interpreted and analyzed in terms of a superconductor-to-insulator quantum phase transition\cite{hebpal}.  This picture requires the existence of a vortex solid phase that persists up to a critical field $H_c$ that in some experiments is a substantial fraction of the superconducting upper critical field, $H_{c2}$\cite{yazkap}.   In contrast, other experiments suggest that quantum fluctuations associated with the small superfluid density destroy the superconducting state at fields below $H_c$.  In previous work, we proposed that quantum fluctuations cause the vortex solid to melt below $H_c$ to form a quantum vortex liquid\cite{chervenak1}.  A number of other studies also substantiate the claim that 2D vortices are susceptible to disorder induced quantum fluctuations in single-layer\cite{qvllayer} and multilayer systems\cite{qvlmulti}.  In this work, we present strong evidence of a novel regime, $R_N\geq R_Q$, in which films which are zero-field superconductors do not superconduct in any finite magnetic field.  We propose that strong mesoscopic and quantum fluctuation effects prevent the formation of the vortex solid phase.  

The Bi/Sb films used in these experiments were deposited on a fire-polished glass substrate held near 4 K on the cold stage of a dilution refrigerator \cite{hsuthesis,strongin}.  The film resistance was measured in a 2 mm x 3.66 mm region with the standard four terminal technique using a lock-in amplifier at low frequency.  The voltage in the sample was measured in the regime which was linear with excitation in the film and, for small sample voltage, checked against the slope of DC current-voltage characteristics at zero bias.  A neighboring region of film was checked for uniformity of $T_{c0}$ and $R_N$ in each sample.  The reported values for applied fields include a correction for flux trapping in the superconducting solenoid.

An example of the TAFF signature exhibited by many two dimensional systems is shown in Fig. 1a.  The tails of the resistive transitions of this Bi/Sb film ($R_N/R_Q =$ .86, $T_{c0} =$ .81 K) follow $R=R_0 exp(-T_0/T)$ (where $R_0$ is a roughly field independent prefactor\cite{chervenak1}) over the range  0.05 $< H/H_{c2} <$ 0.5 and 0.06 $< T/T_{c0} <$ 0.5.   The activation energy, $T_0$ depends on magnetic field, $T_{c0}$, and film thickness, $t$, as 
\begin{mathletters}
\begin{equation} T_0 \propto T_{c0}t log {{H_0}\over H}
\end{equation}
\end{mathletters}
where $H_0$ is the field at which the activation energy extrapolates to zero.  The inset of Fig. 1a demonstrates the logarithmic field dependence of $T_0$ measured for the data in Fig. 1a.  We have observed similar behavior in all Bi/Sb films in the range 0.6 $<T_{c0}<$ 3 K.  This TAFF behavior agrees with models that ascribe the dissipation to the thermally activated motion of defects in a collectively pinned, glassy, vortex solid.  Those models predict that a film superconducts in the low temperature limit, as long as $H<H_0$, i.e. there is a finite barrier to dissipative processes.  In earlier work, we pointed out that $H_0<H_{c2}$ in Bi/Sb films, implying the existence of a regime, $H_0<H<H_{c2}$, which neither superconducts nor insulates.  We referred to it as a Quantum Vortex Liquid Regime (QVL).  Here, we present evidence that in more strongly disordered films, $H_0\rightarrow$ 0 and the QVL persists to arbitrarily low magnetic field and temperature.  

We have systematically varied disorder in a series of Bi/Sb films to yield the dependence of the characteristic field, $H_0$, on $T_{c0}$ shown in Fig. 2.  The solid line is a linear fit to the data with a slope of 1.25 T/K and an x-axis intercept of $T_{c0}*\approx$0.4 K.  This finite intercept is significant as it implies that the average barrier to the thermal activation of dissipative processes is zero in films with $T_{c0}<T_{c0}*$.  Thus, TAFF behavior should cease when $T_{c0}<T_{c0}*$ or, according to the inset of Fig. 2, $R_N/R_Q\approx$ 1.  Indeed, as shown in Fig. 1b, the low field transport of Bi/Sb films with $T_{c0}<T_{c0}*$ deviates from TAFF behavior.  These data come from a Bi/Sb film with $R_N/R_Q =$ 1.17 and a $T_{c0}=$0.255 K that was deposited in the same experiment as the film in Fig. 1a.  The tails of its transitions continuously curve away from a simple activated form and tend to level off, with the slope of each curve asymptotically approaching zero with decreasing temperature (i.e., approaching metallic behavior).  We were unable to obtain good fits to these data using any simple power-law relationship (i.e. $log(R)\propto T^a(log(H))^b$ where a and b are constants) that have been suggested for vortex solids with non-logarithmic interactions\cite{fish1}.  Non-thermally activated resistive transitions have been observed for all applied fields in two Bi/Sb films with $R_N$ and $T_{c0}$ of 4.85 k $\Omega$ and .255 K and 4.5 k $\Omega$ and .336 K respectively \cite{other flattening}. 

The lack of TAFF at any field implies the absence of a zero resistance, vortex solid phase even at T=0.  Furthermore, the systematic evolution of $H_0$ with $T_{c0}$ and the coincidence of $T_{c0}*$ with the disappearance of TAFF strongly suggests that a QVL replaces the vortex solid for $T_{c0}<T_{c0}*$.  This coincidence makes unlikely the alternative explanation that with increasing disorder the melting temperature of the solid phase becomes lower than accessible temperatures.  The replacement of the vortex solid by the QVL that we propose, however, may not be complete.  Patches of liquid and solid with a range of melting temperatures could coexist in this regime and render the same behavior.  These patches may arise naturally (see discussion below) in a uniformly disordered system.  The possibility that macroscopic inhomogeneities have occurred in the low $T_{c0}$ films is ruled out by the recovery of TAFF with a submonolayer deposition of Bi on top of the film (See Fig. 1a). 

Recent theories predict that nonthermal quantum and/or mesoscopic fluctuation effects can be sufficient to melt or partially melt the vortex solid when $R_N\approx R_Q$ in support of this interpretation.  Blatter and Ivlev proposed that quantum fluctuations make vortices "fuzzy" or their cores effectively larger\cite{blatter}.  This effect causes the vortex solid to melt at lower fields than strictly thermal models would predict\cite{feigelman}.  Melting at zero temperature\cite{kes,chervenak1} occurs at a field given by,
\begin{mathletters}
\begin{equation} {{H_m}\over{H_{c2}}}=1-1.2exp{{\pi^3{c_L}^2R_Q}\over {R_N}}
\end{equation}
\end{mathletters}
For a reasonable Lindemann parameter $c_L \sim$ .2, Eq. 2 predicts that quantum fluctuations suppress $H_m$ to near zero for $R_N \approx R_Q$.  Within this framework, the onset of non-activated transport below $T_{c0}*$ occurs when quantum fluctuations makes the average effective size of the vortex cores so large that $H_m$ becomes zero.  

Mesoscopic fluctuations in the number of electronic states near the Fermi energy may also contribute to or be the primary cause of the rise of the QVL.  As noted previously, the number of electronic states in a superconducting coherence volume that are within $\Delta$, the superconducting energy gap, of the Fermi energy approaches two in the limit $R_N\simeq R_Q$\cite{hsu,chervenak2}.  Fluctuations in this number will be of the same order and will manifest themselves as spatial fluctuations in the order parameter amplitude, or $H_{c2}$\cite{zhou} or $T_{c0}$\cite{ghosal,belkirk}.  Recently, we observed structure in zero magnetic field DC-IV curves of Bi/Sb films with $T_{c0} < T_{c0}*$ that was consistent with the presence of fluctuations in the order parameter amplitude\cite{chervenak2}.  Spivak and Zhou showed that in finite magnetic field mesoscopic fluctuations lead to a distribution of $H_{c2}$s with a width given by $<(\delta H_{c2})^2>/({H^0}_{c2})^2=\gamma(R_N/R_Q)^2$ in a single sample ($\gamma \sim 1$)\cite{zhou}.  To account for our observations we propose that areas with lower than average $H_{c2}$ and larger than average vortices grow and create more of the measured dissipation with increasing $R_N$.  The measured activation energy must then be an average over the distribution of energies.  On mesoscopic scales, the energy to activate defects in the vortex solid or to unpin vortices in these low $H_{c2}$ regions can be much smaller than the measured activation barrier.  The deviations from activated behavior occur when regions with $H_{c2}\approx$ 0 percolate across the film.  

Our data provide evidence, in accord with the above theories, that the effective size of the vortex cores grows anomalously fast as $R_N\rightarrow R_Q$.  The evidence comes from interpreting the field scale $H_0$.  Within the TAFF models, $H_0$ (cf. Eq. 2) should be proportional to $H_{c2}$.  The latter sets the characteristic length scale, ${\xi} = \sqrt{\Phi_0/2\pi H_{c2}}$, the vortex core radius, in the logarithmic vortex-vortex interaction potential, where $\Phi_0$ is the superconducting flux quantum.  In accord with this expectation, the linear dependence of $H_0$ on $T_{c0}$ shown in Fig. 2 quantitatively agrees with the mean field, dirty limit, dependence of $H_{c2}$ on $T_{c0}$ (i.e. $H_{c2}={{\Phi_0k_BT_{c0}}\over{.18hD}}$, where $D$ is the electronic diffusivity).  However, the finite x-axis intercept in Fig. 2 implies that $H_0<H_{c2}$ and correspondingly, the effective size of the vortices in the Bi/Sb films, ${\xi_q=\sqrt{\Phi_0/2\pi H_{0}}}$ exceeds the superconducting coherence length as $T_{c0}\rightarrow T_{c0}*$.  This discussion leads to the picture that the effective vortex core size relevant to vortex-vortex interactions grows faster with decreasing $T_{c0}$ than expected for a mean field dirty limit superconductor.

The absence of a vortex solid phase precludes the existence of a magnetic field tuned superconductor-to-insulator quantum phase transition.  It implies the existence of an intermediate metallic regime such as the QVL.  In accord with this assertion, the magnetotransport in Bi films at higher magnetic fields do not follow the scaling behavior that has been taken as the primary evidence of a direct SIT.  Figures 3a and b show the magnetotransport of two films, one with one with $T_{c0}>T_{c0}*$ and one with $T_{c0}<T_{c0}*$ in this regime.  It is not possible to identify, in either data set, the single, critical magnetic field for which $dR/dT = $0 in the low temperature limit necessary for scaling the data.  Instances for which scaling "works" are either at higher temperatures or in lower $R_N$ films whose normal states have a weaker temperature dependence than shown in Fig. 3 (with notable exceptions \cite{hebpal}).  To fix problems with scaling like those in Fig 3a, Gantmakher suggested that the critical field corresponds to where the 2nd derivative of the resistance with respect to temperature is zero rather than the first derivative \cite{gantmakher}.  Until theory justifies this modification it seems more reasonable to assume that no critical field exists.  Within the quantum vortex liquid interpretation, the transition to the normal insulating state at high magnetic fields would be expected to be a crossover rather than a phase transition.  The crossover would be similar to that which occurs near $H_{c2}$ at non-zero temperatures in high-$T_c$ superconductors \cite{high TC}.  The smooth evolution of the data in Figs. 3a and b seem more consistent with such a crossover.  

In summary, thermally activated flux flow observed in superconducting Bi/Sb films in applied magnetic fields disappears when film disorder approaches a normal state sheet resistance $R_N = R_Q = 4.12 k\Omega$.  The crossover in homogeneously disordered Bi/Sb coincides with the point at which the activation barrier observed in the less disordered films extrapolates to zero. The quantum vortex liquid regime (QVL) arises well before $T_{c0}$ is suppressed to zero by disorder and exhibits finite resistance in any magnetic field to ultralow temperatures.  The presence of the QVL strongly suggests that disorder induces either mesoscopic or quantum fluctuations in the superconducting state which ultimately become strong enough to determine the electronic phase. 

\acknowledgments
We wish to acknowledge the support of NSF grants DMR-980193 and DMR-9502920, helpful conversations with Fei Zhou, Brad Marston, Boris Spivak, and Nandini Trivedi.  We are grateful to Sean Ling for a critical reading of the manuscript.

\begin{figure}
\caption{ (a)  Thermally activated dissipation in Bi/Sb with $R_N$ = 3.53 k $\Omega$ and $T_{c0} = $.81 K in applied fields of 0 T, .01 T, .02 T, .05 T, .1 T, and .2 T. The solid lines are the fit to the activated portions. Similar behavior is observed in Bi/Sb films with .6 $< T_{c0} < $3 K.  The inset shows the logarithmic dependence of the activation barrier $T_0$ on applied field.  The x-axis intercept of the fit is defined as $H_0$, the characteristic field of the activation barrier.  (b)  Non-thermally activated dissipation in Bi/Sb film with $R_N = $4.84 k $\Omega$ and $T_{c0} = $.255 K in applied fields: 0 T, .0002 T, .002 T, .003 T, .007 T, .011 T, .022 T, .05 T, .08 T.}
\label{autonum}
\end{figure}

\begin{figure}
\caption {The field $H_0$ decreases linearly with $T_0$ for Bi/Sb films.  The solid line shows a fit to the data which extrapolates to and implies that $H_0$ is zero for Bi/Sb films with $T_{c0}\approx$ .4 K.}  
\label{autonum}
\end{figure}

\begin{figure}
\caption {Transport in Bi/Sb films with $R_N$ above and below $R_Q$ in fields ranging from zero tesla to near $H_{c2}$.  Neither data set features a 'critical field' $H_c$ at which the resistive transition exhibits $dR(H_c)/dT =$ 0 over a reasonable range of temperature. (a) Bi/Sb with $T_{c0} =$ 1.44 K in fields 1.4 T, 1.45 T, 1.5 T, 1.55 T, 1.6 T, and 1.8 T.  Inset: Resistive transitions of the same film in applied fields 0 T, .1 T, .2 T, .5 T, 1.4 T, 1.55 T, 3 T (b)  Transport in a Bi/Sb film with $T_{c0} = $ .255 K in fields .04 T, .1 T, .15 T, .2 T, .36 T.  Inset: Resistive transitions of the same film in applied fields 0 T. 0.02 T, 0.1 T, 0.2 T, 0.5 T}
\label{autonum}
\end{figure}


\begin{references}

\bibitem[*]{byline}  Also at National Institute of Standards and Technology, Boulder, CO. E-mail: jamesch@boulder.nist.gov.

\bibitem{dsfish} D. S. Fisher, Phys.Rev. Lett. {\bf 78}, 1964 (1997).

\bibitem{giam} T. Giamarchi and P. Le Doussal, Phys. Rev. B. {\bf 55}, 6577 (1997).

\bibitem{blatter} G. Blatter and B. Ivlev, Phys. Rev. Lett {\bf 70}, 2621 (1993); G. Blatter, et al. {\bf 50}, 13013 (1994).

\bibitem{ghosal}  A. Ghosal, M. Randeria, and N. Trivedi,  Phys. Rev. Lett.  {\bf 81}, 3940 (1998).

\bibitem{shim} E. Shimshoni, A. Auerbach, and A. Kapitulnik.  Phys. Rev. Lett. {\bf 80}, 3352 (1998).

\bibitem{ephron}  D. Ephron, A. Yazdani, A. Kapitulnik, and M. R. Beasley, Phys. Rev. Lett. {\bf 76}, 1529 (1996); N. Mason and A. Kapitulnik, Phys. Rev. Lett.  {\bf 82}, 5341 (1999).

\bibitem{kes1} M. H. Theunissen, B. Becker, and P. H.  Kes. {\it private communication}.

\bibitem{fisher} M. P. A. Fisher, G. Grinstein, and S. Girvin,  Phys. Rev. Lett. {\bf 64}, 587 (1990).

\bibitem{belkirk} T. R. Kirkpatrick and D. Belitz,  Phys. Rev. Lett. {\bf 79}, 3042 (1997).

\bibitem{sitreviews}  A. F. Hebard in Strongly Correlated Electronic Materials, ed. K. Bedell.  Addison-Wesley, Reading, MA (1993);  N. Markovic and A. M. Goldman, Physics Today {\bf 51}, 39 (1998).

\bibitem{sondhi} S. L. Sondhi, S. M. Girvin, J. P. Carini, and D. Shahar, Rev. Mod. Phys. {\bf 69}, 315 (1997).

\bibitem{larkin} A. I. Larkin and O. Ovchinnikov, J. Low. Temp. Phys. {\bf 34}, 409 (1979).

\bibitem{zhou} B. Spivak and F. Zhou,  Phys. Rev. Lett. {\bf 74}, 2800 (1995).

\bibitem{kes} P. Berghius and P.H. Kes, Phys. Rev. B. {\bf 47}, 5126 (1993).

\bibitem{feigelman}  M. V. Feigelmann, V. B. Geshkenbein, and A. I. Larkin, Physica C {\bf 167}, 177 (1990);  H. J. Jensen, P. Minnhagen, E. Sonin, and M. Weber, Europhys. Lett. {\bf 20}, 463 (1992).

\bibitem{TAFF1} for example, W. White, M. R. Beasley, and A. Kapitulnik, Phys. Rev. B. {\bf 49}, R7084 (1994).

\bibitem{chervenak1}  J. A. Chervenak and J. M. Valles, Jr. Phys. Rev. B. {\bf 54}, R15649 (1996).

\bibitem{liu} D. Haviland, Y. Liu, and A. M. Goldman, Phys. Rev. Lett. {\bf 62}, 2180 (1989).
  
\bibitem{hsu} S.-Y. Hsu, J. A. Chervenak, and J. M. Valles, Jr. Phys. Rev. Lett. {\bf 75}, 132 (1995).

\bibitem{chervenak2}  J. A. Chervenak and J. M. Valles, Jr.  Phys. Rev. B. {\bf 59}, 11209 (1999).

\bibitem{hebpal} M. Paalanen and A. F. Hebard,  Phys. Rev. Lett. {\bf 65}, 927 (1990).

\bibitem{yazkap} for example, A. Yazdani and A. Kapitulink,  Phys. Rev. Lett. {\bf 74}, 3037 (1995).

\bibitem{qvllayer} T. Sasaki, et al. Phys. Rev. B {\bf 57}, 10889 (1998).

\bibitem{qvlmulti} C. Attanasio, et al. Phys. Rev. B. {\bf 53}, 1087 (1996);  N. Y. Fogel, V. G. Cherasova, O. A. Koretzkaya, and A. S. Sidorenko, Phys. Rev. B. {\bf 55}, 85 (1997).

\bibitem{hsuthesis} S.-Y. Hsu,  Ph. D. thesis. Brown University (1995).

\bibitem{strongin} M. Strongin, R. Thompson, O. Kammerer, and J. Crowe,  Phys. Rev. {\bf B1}, 1078, (1970).

\bibitem{fish1}  M. P. A. Fisher, T. A. Tokuyasu, and A. P. Young,  Phys. Rev Lett. {\bf 66}, 2931 (1991);  D. S. Fisher, M. P. A. Fisher, and D. A. Huse, Phys. Rev. B. {\bf 43}, 130 (1991). 

\bibitem{other flattening} The character of the deviations from activated behavior shown in Fig. 1b differ qualitatively from those observed previously in other, less disordered, thin film systems.  In particular, Ephron and coworkers \cite{ephron} reported that the resistance of MoGe films ($R_N \sim$ 1.5 k $\Omega$) has an activated temperature dependence that abruptly "flattens", i.e. becomes temperature independent.  The deviations from activated behavior shown in Fig. 1b are significantly more gradual and only appear in films with $R_N$ approximately equal to $R_Q$. Moreover, Bi/Sb films with sheet resistances that are comparable to those for flattening in MoGe films show only activated behavior to temperatures and resistances far below where the flattening occurs in MoGe.  We do not understand the source of these qualitative differences in the data.
    
\bibitem{high TC} K. Karpinska, et al. Phys. Rev. Lett. {\bf 77}, 3033 (1996).

\bibitem{gantmakher} V. F. Gantmakher, M. Golubkov, V. J. Dolgopolov, G. E. Tsydynzhopov, and A. A. Shashkin, JETP Lett. {\bf 68}, 363 (1998).

\end{references}
\end{document}